# Small-body deflection techniques using spacecraft: techniques in simulating the fate of ejecta

Stephen R. Schwartz[a,*], Yang Yu[a], Patrick Michel[a], and Martin Jutzi[b]

[a]*Labratoire Lagrange, Univ. Nice, CNRS, Observatoire de la Côte d'Azur, Boulevard de l'Observatoire, CS 34229, 06304 Nice Cedex 4, France*
[b]*University of Bern, Center for Space and Habitablity, Physics Institute, Sidlerstrasse 5, 3012 Bern, Switzerland*

**Abstract**

We define a set of procedures to numerically study the fate of ejecta produced by the impact of an artificial projectile with the aim of deflecting an asteroid. Here we develop a simplified, idealized model of impact conditions that can be adapted to fit the details of specific deflection-test scenarios, such as what is being proposed for the AIDA project. Ongoing studies based upon the methodology described here can be used to inform observational strategies and safety conditions for an observing spacecraft. To account for ejecta evolution, the numerical strategies we are employing are varied and include a large *N*-Body component, a smoothed-particle hydrodynamics (SPH) component, and an application of impactor scaling laws. Simulations that use SPH-derived initial conditions show high-speed ejecta escaping at low angles of inclination, and very slowly moving ejecta lofting off the surface at higher inclination angles, some of which reimpacts the small-body surface. We are currently investigating the realism of this and other models' behaviors. Next steps will include the addition of solar perturbations to the model and applying the protocol developed here directly to specific potential mission concepts such as the proposed AIDA scenario.

| Nomenclature | |
|---|---|
| $U$ | relative impact velocity (m/s) |
| $u$ | speed of grain/ejecta fragment |
| $R$ | crater radius (m) |
| $C_1$ | constant (overall linear velocity dependence) |
| $a$ | projectile size (m) |
| $\rho$ | target bulk density (kg/m$^3$) |
| $\delta$ | projectile bulk density (kg/m$^3$) |
| $\mu$ | constant (exponential decay) |
| $\nu$ | constant (exponential density dependence) |
| $n_1$ | constant (smallest initial distance of interest) |
| $n_2$ | constant (largest initial distance of interest) |
| $p$ | constant (exponential cut-off) |
| $r_c$ | radius of curvature of boundary shapes used in simulations |
| $\phi$ | opening angle of spherical boundary shapes used in simulations |

\* Corresponding author. Tel.: +33-4-92-00-31-56; fax: +33-4-92-00-30-58.
*E-mail address:* srs@oca.eu.

# 1. Introduction

It has been estimated that every year, asteroids measuring 4 m across enter Earth's atmosphere and detonate (Collins et al., 2005), and that asteroids of 100 m tend to impact the Earth every 5,000 years. On average, collisions with 4-kilometer asteroids are occurring about every 13 million years (Harris, 2009). In 2005, a United States Congressional mandate called for NASA to detect, by 2020, 90% of near-Earth objects (NEOs) with diameters of 140 m or greater (Space Sciences Board, 2009).

As sky surveys are performed, and detection strategies are developed for discovering small bodies that may be on trajectories to collide with the Earth, it is prudent to concurrently develop and refine mitigation strategies, including those that could, as needed, alter the paths of such hazardous objects. However, any discussion of efforts to deviate the path of a small Solar System body must respect the wide variety of shapes, sizes, densities, porosities, and chemical makeups of the asteroids that lie in Earth-crossing orbits. As a by-product, mitigation tests would provide answers to important scientific questions regarding the physical properties of near-Earth asteroids, and their response to a deflection technique.

One of the mitigation strategies explored by the United States' National Research Council's (NRC) "Committee to Review Near-Earth Object Surveys and Mitigation Strategies" involves using an impactor spacecraft to deflect an NEO by crashing into it at speeds of up to 10 km/s or more (Space Sciences Board, 2010). Such a strategy had already been considered by ESA in its Phase-A study of the Don Quixote mission recommended by its Near-Earth Object Advisory Panel (NEOMAP; Harris et al., 2006). This concept included an artificial impactor and an orbiter to characterize the asteroid target and measure the deflection resulting from the impact. The NEOShield Project aims to design a general NEO defense strategy based upon momentum transfer via kinetic impact (Harris et al., 2013). Begun in 2012, the NEOShield Project is being funded for 3.5 years by the European Commission in its FP7 program. It is, primarily, but not exclusively, a European consortium of research institutions that aims to analyze promising mitigation options and provide solutions to the critical scientific and technical obstacles involved in confronting threats posed by small bodies that cross Earth's orbit. The Asteroid Impact Deflection and Assessment (AIDA) mission is the first actual deflection test currently under study by two main space agencies (Michel et al., 2015). Comprising AIDA are the NASA Double Asteroid Redirection Test (DART) mission, which began its Phase-A study in October 2015, and the ESA Asteroid Impact Monitor (AIM) rendezvous mission, which began its Phase-A/B1 study in March 2015. DART is being devised to test our ability to deflect an asteroid using a kinetic impactor, while the AIM mission seeks to characterize the DART target—the binary near-Earth asteroid (65803) Didymos—before, during, and after the impact (fall 2022). In order to determine its observational strategy and positioning during and after the DART impact, the AIM component of the mission requires an educated assessment of how the ejecta produced by a kinetic impactor may evolve. AIDA, if funded for launch in 2020 and arrival to the target in 2022, will be the first fully documented impact experiment at real asteroid scale, allowing numerical codes to be tested and used for similar and other scientific applications at those scales.

The first large-scale, active impact experiment on a small body was performed on July 4, 2005, by the Deep Impact mission (NASA). The target of Deep Impact was Comet 9P/Tempel 1 and the impact excavated material from well below its surface, allowing its investigation both by the Deep Impact spacecraft during its comet flyby and by remote sensing from Earth and from elsewhere in the Solar System (e.g. Rosetta, Spitzer Space Observatory). As recalled by A'Hearn & Combie (2007), the primary goal of Deep Impact was to understand the difference between the surface of a cometary nucleus and its interior and thus to understand how the material released spontaneously by the nucleus is related to the primitive volatiles that were present in the protoplanetary disk. The other important goal was to understand the physical properties of the outer layers of the comet. Therefore, the goal was not to make a deflection test and with a projectile's mass of about 370 kg and an impact speed about 10 km/s, given the comet's diameter of 6 km, the impact energy was too small to lead to any measurable deflection of the comet's trajectory. The mission was a success as the impact occurred as predicted and plenty of data on the comet were obtained. Regarding the outcome of the impact itself, as reported by Holsapple & Housen (2007), conventional cratering may not be the sole key to the observed plume of the Deep Impact event. These authors argue that acceleration mechanisms from vaporizing ices or perhaps solar pressure most likely moved the material more quickly to the 100-km ranges. In fact, a volatile subsurface could greatly enhance the amounts of ejected mass and the uncertainty of those "comet" mechanisms makes it difficult to relate the observed plume mass to the crater using measured cratering events. Richardson et al. (2007) developed a first-order, three-dimensional, forward model of the ejecta plume behavior of the Deep Impact experiment and adjusted the model parameters to match the flyby-spacecraft observations of the actual ejecta plume, image by image. Individual projectiles (ejecta particles) in flight could not be observed and the collective behavior of the ejecta particles was rather monitored while they formed a hollow, expanding, cone-shaped cloud. The early stages of the crater excavation flow have been modeled using the Maxwell Z-model developed by Maxwell & Seifert (1974) for explosion craters and then extended to impact craters by Maxwell (1977). Because of its underlying assumptions, the Z-model represents only a good first-order description of cratering excavation flow and lacks many of the finer variations present in even a simple cratering event. Richardson et al. (2007) often refer to it but use scaling relations and numerical computation to follow the particle behaviors at the point of launch. By matching the outcome to the observations, they could derive an estimate of the strength that is in some dispute and the bulk density of the comet. However, they recognized that the large-end error in the bulk density estimate is due to uncertainties in the magnitude of coma gas pressure effects on the ejecta particles in flight, emphasizing again the non-traditional aspect of the cratering mechanics of this event.

A first experiment on an asteroid will be performed by the Small Carry-on Impactor (SCI) onboard the Japanese sample return space mission Hayabusa-2 successfully launched on December 3, 2014 (Arakawa et al., 2013). In this case again, the aim is not to perform a deflection test of the 750-m-diameter target (162173) Ryugu. The impactor's mass is only about 2 kg and its impact speed about 2 km/s. The goal is rather to reveal the subsurface properties of the asteroid and to possibly take a sample inside the crater. Nevertheless, it will allow us to observe a possibly traditional cratering event on an asteroid surface for the first time and check scaling laws and numerical models at asteroid scale.

Fahnestock & Chesley (2013) studied the dynamical behavior of ejecta produced by the ISIS Kinetic Impactor demonstration proposed but finally not selected to accompany the OSIRIS-Rex (NASA) sample return mission on the asteroid Bennu (Lauretta et al., 2015). They studied the dynamics of the ejecta applying scaling laws and then following their trajectories under the influence of the asteroid, for which they used a shape-model-derived full polyhedral body gravity, the differential solar tides acceleration, as well as the differential solar radiation pressure acceleration, accounting for potential shadowing and realistic particle optical properties. Using the impact conditions planed for these events and assuming strength properties for the asteroid, they computed the evolutions of particle tracers and could determine regions with higher impact probabilities of high-velocity ejecta with a spacecraft. Richardson and Taylor (2015) investigated the fate of ejecta in the (66391) 1999 $KW_4$ binary asteroid system also using the model described by Richardson et al. (2007) for the ejecta plume and following tracer particles accounting from the gravitational attraction of the two binary components. However, these first simulations need to be expanded, and the next step is to perform such modeling using higher fidelity simulations of the cratering event and of the ejecta evolutions, accounting for the large number of individual particles that form the ejecta cloud as well as their potential mutual interactions in addition to other forces acting on them.

In this study, we work to establish different methodologies that can be used to set up the initial state of the excavation resulting form a kinetic impactor in order to follow the ejecta evolution and fate. This study and the methodologies developed here will be of use in future studies to specific scenarios, such as the kinetic impactor involved in the AIDA mission, and other mission concepts involving kinetic impactors into small bodies. In Section 2, we describe our different approaches to account for initial ejecta velocities and our methods of following ejecta evolution based on various perturbations that influence their trajectories. In Section 3, we provide the outputs generated based on our example impact scenario using these ejecta models. In Section 4, we put our study in a broader context and discuss the ongoing and future progression of our work.

## 2. Numerical Methodology

Here we present the different approaches we use to set up the gravity field and assign ejecta sizes, positions, and velocities based on the post-fragmentation phase of a kinetic impactor strike. Once these properties are defined, we use the *N*-Body code PKDGRAV (Stadel, 2001) to compute their trajectories, accounting for up to tens of millions of ejecta particles.. As a first step, we focus on ejecta with sizes greater than a few centimeters and do not consider the effect of solar radiation pressure or the possible perturbations due to solar (and planetary) tides to compute their evolutions. A thorough account of these effects and their relevance are necessary when applied to specific impact scenarios involving specific targets, but do not change the conclusions derived from this study.

To define the ejecta properties, we distinguish between three very different approaches to modeling the state of the system immediately following the hypervelocity impact phase. Each approach is based on contemporary understandings of the impact physics, either through theoretical works based on empirically derived scaling relationships, or through direct numerical simulations of the fragmentation phase itself. These approaches have their respective positive and negative attributes (see Section 2.4 and Section 4), and each fall into unique categories: (1) *analytical scaling laws*, empirically derived from experiments, that describe ejecta properties as they are launched from the impact site, (2) *hydrocodes* (grid codes, smoothed-particle hydrodynamics [SPH] codes) to compute the impact and fragmentation phase, and (3) the use of *N-Body simulations* to compute the whole impact process in order to investigate more precisely the very low-speed material lofted from the surface. In the first case, we apply analytical scaling laws to our simple deflection scenario to compute ejection velocities, take these as inputs to our *N*-Body code, and then compute the fate of ejecta using full gravity. In the second case, we use a numerical code capable of modeling the impact event, then, once the impact phase is complete, we take the ejecta properties as inputs to our *N*-Body code and compute the fate of ejecta using full gravity and soft-sphere collisions between ejected particles. In the third case, an *N*-Body approach is used to solve for the state of the low-speed, post-impact phase ejecta, with validation provided by comparison with scaling laws, before modeling the ejecta evolution. In this paper, we detail just the first two approaches. The first of these comprises the use of the scaling laws of Housen & Holsapple (2011) to assign the position and velocity distributions to ejecta (Section 2.1), while the second approach comprises the use of the SPH simulations of Jutzi & Michel (2014) to compute the impact (Section 2.2). In addition, we are pursuing the use of other hydrocodes and scaling-relation models of ejecta (see Section 4).

A hypervelocity impact will produce ejecta fragments composed of high-speed particles that will quickly escape the system. However, after this time, the primary concern for an observing spacecraft is the material that is lofted off the asteroid surface that may linger in its vicinity. It is not the high-speed particles that escape initially, but this low- to moderate-speed material that poses a more-difficult-to-assess threat to operations of visiting spacecraft. Low-speed fragments may not immediately escape the system, but rather go into quasi-orbit or trace slow, complex paths within the system before eventually escaping or falling back onto the asteroid's surface [this assertion is supported by studies reported by, e.g., Fahnestock & Chesley, S.R. (2013); Yu et al. (2015); Richardson & Taylor (2015); as well as our own preliminary results from a kinetic impactor study involving the asteroid (25143) Itokawa (Drube et al., 2015)]. In order to build an informative model to assess probable locations of hazardous debris accounting for all ejection velocities, each method we discuss depends on allowing discrete pieces of ejecta to evolve in an *N*-Body code.

An *N*-Body code is well suited to model discrete material in the subsonic regime and to account for the gravitational environment. We use PKDGRAV, a parallel gravity tree code that computes collisions between spherical particles (Richardson et al., 2000), which has been successfully used to reproduce formation scenarios of asteroid families (e.g., Michel et al., 2001) and to reproduce laboratory experiments on granular materials under various regimes and contexts (e.g., Matsumura et al., 2014; Schwartz et al., 2014). The tree component of the code provides a convenient means of consolidating forces exerted by distant particles, reducing the computational cost, while the parallel component divides the work evenly among available processors, adjusting the load each time step according to the amount of work done during the previous force calculation. A straightforward second-order leapfrog scheme is used for the integration and to compute gravity moments from tree cells to hexadecapole order (see Stadel, 2001). Particles are considered to be finite-sized spheres and contacts are identified each step using a fast neighbor-search algorithm also used to build the tree. The code was adapted to treat hard-sphere collisions for planetesimal modeling (Richardson et al., 2000) and later for granular material modeling (Richardson et al., 2011; Schwartz et al., 2012). At our disposal are two discrete element method (DEM) collisional routines: the soft-sphere discrete element method (SSDEM), based upon the work of Cundall & Strack (1979) and outlined in Schwartz et al. (2012), and the hard-sphere discrete element method (HSDEM), outlined in Richardson et al. (2000) and Richardson et al. (2011). The choice of collisional routine will depend upon particle sizes and densities, typical collisional speeds, and other more nuanced factors. In dense granular regimes, when contacts are sufficiently complex—that is, when contacts are long lasting, requiring explicit treatment of frictional forces, or when the propagation speeds of disturbances (material sound speeds) are important—SSDEM is the correct choice. When the system is sparse and collisions are less frequent, HSDEM is usually the best choice for its superior computational speed in such regimes. Given the importance of granular interactions between ejecta particles, in the study thus far we use exclusively SSDEM, but as mentioned in Section 4, there may be a role for HSDEM in evolving large numbers of late-stage ejecta particles (i.e., when the system is sparse and physical interactions between grains are relatively few).

We define a baseline setup to illustrate and test our methodologies. This setup includes a sphere of diameter 150 m as the target, which is of of a size commensurate with the NEOShield specifications as well as with the radar observations of the secondary of the Didymos system (performed in 2003, Lance Benner—personal communication). We take a bulk density of 1,300 kg/m$^3$, and impactor mass and density of 400 kg and 1,000 kg/m$^3$, respectively, to be consistent with the suite of SPH impact simulations that we use, performed by Jutzi & Michel (2014).

In this section, we first explain our approaches to define the post-impact ejecta properties that use experimentally derived scaling relations (Section 2.1) or SPH simulations (Section 2.2) before showing how each are then used in an *N*-Body code to compute the ejecta fate (Section 2.3). Lastly in this section, we discuss some of the current limitations of these approaches (Section 2.4).

*2.1. Initial impact conditions using empirically derived scaling laws*

Comparisons of scaling laws and impact simulations show commensurate results for impact velocities larger than 5 km/s (Jutzi & Michel, 2014). This lends credence to testing an approach that solves for the ejecta conditions at the end of the fragmentation phase based upon scaling laws alone. Empirically derived scaling relationships have been developed by, e.g., Chabai (1965); Holsapple & Schmidt (1982); Housen & Holsapple (2011). Some of these teams examined the data from specific experiments and assessed how the ejecta velocity and mass distributions depend on the conditions of an impact event, in particular on the impact speed and target properties such as strength and porosity. Fitting the data for several classes of materials, this contributed to point-source scaling models for the ejecta mass and velocity distributions.

We use these results to assign ejecta mass and velocity distributions, assuming a crater size of 27 m across (of the same order as what is expected from these scaling laws and simulations). We do not include collisional interactions between ejecta particles, but do include gravitational interactions between them. Also, all ejected particles are assigned their initial velocities at the start of the simulation, approximating the excavation time to be instantaneous (this can be mitigated if we determine that it is important to determining the ejecta fate). Our aim in this study is not to make a precise prediction of the actual event with these initial conditions, but rather to explicate and demonstrate our methods. Although we use a highly idealized scenario, we describe here our approach to implement the chosen conditions using specific values for the relevant parameters, and then provide the results of this test study in Section 3.

The first step in this approach is to numerically simulate the filling of a *tub* with particles; we use sizes 14–20 cm from a power-law distribution of diameter with slope $-2.5$ (a slope within the range of what the Hayabusa mission observed on Itokawa, as reported by, e.g., Michikami et al., 2008) and integrate until these particles come to rest (arbitrary sizes and distributions can be used to address different regimes). The tub is defined as a spherical shell with radius of curvature $r_{c,tub}$ and "opening angle" $\phi_{tub}$, and placed inside the surface of a larger sphere that represents the asteroid surface (Figs. 1–2). The procedure to simulate the filling of this tub with particles involves computation in several stages using PKDGRAV and wall conditions that are described in-depth by Schwartz, et al. (2012). We skip the details associated with this procedure and simply indicate that the outcome is an ensemble of particles with an initial packing fraction of 63.7% being ejected with an angle of 45° with respect to the local surface tangent and with ejection speeds decreasing with increasing distance from the impact point described by Eq. (14) of Housen & Holsapple (2011):

$$\frac{u}{U} = C_1 \left[\frac{x}{a}\left(\frac{\rho}{\delta}\right)^\nu\right]^{-1/\mu} \left(1 - \frac{x}{n_2 R}\right)^p, \qquad n_1 a \leq x \leq n_2 R. \tag{1}$$

Save for the value of $p$, the parameter values chosen are close to the curve $C5$ shown in Housen & Holsapple (2011), Table 3.: $U = 6000$ m/s, $R = 13.5$ m, $C_1 = 0.55$, $a = 0.457$ m, $\rho = 1300$ kg/m$^3$, $\delta = 1{,}000$ kg/m$^3$, $\mu = 0.41$, $\nu = 0.4$, $n_1 = 1.2$, $n_2 = 1.2$, $p = 3.3$. The tub has a surface radius ($r_{s,tub}$) of $1.3R$; some of the slowest moving particles extend beyond $R$, the nominal crater radius. Although here we use a constant 45°, we are in general free to choose an ejection angle that may vary with distance from the impact point. Note also that the relatively large value of $p$ is used here as a coarse way to account for the presence low-speed, surface-regolith ejecta, which have hardly been addressed empirically to date.

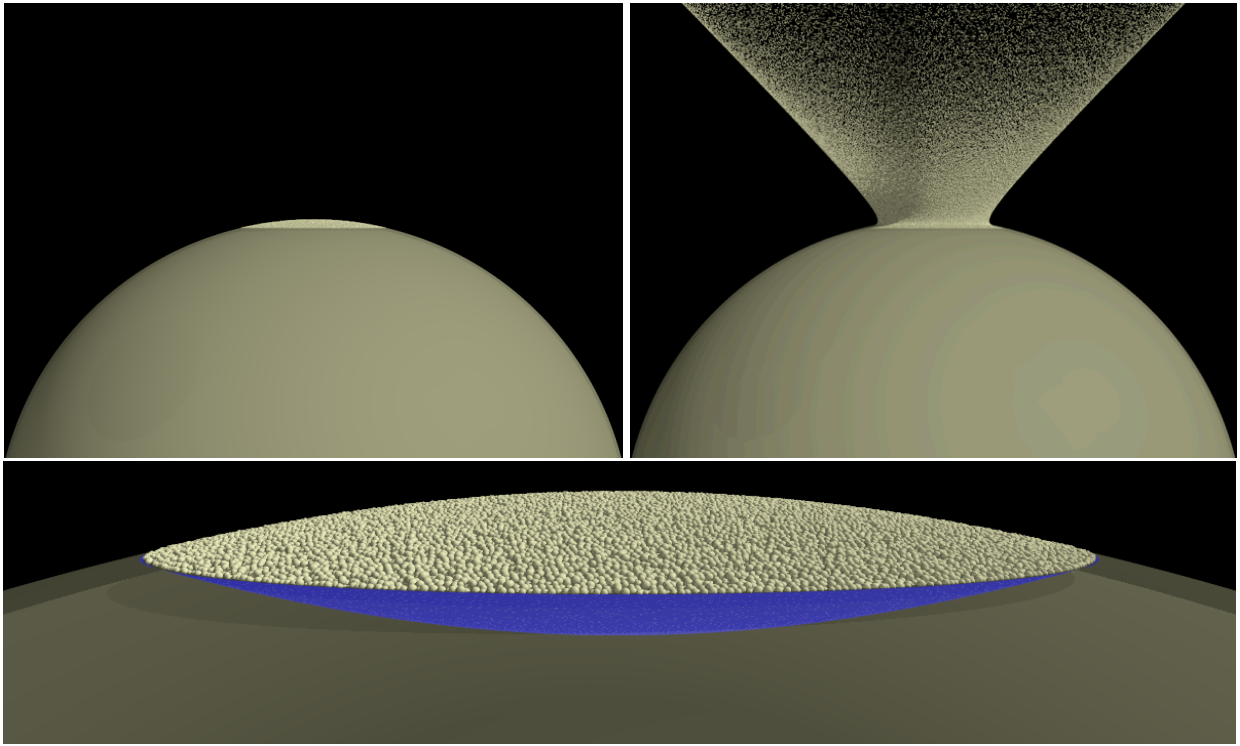

**Figure 1:** The start (top-left frame) of an *N*-body simulation using ejecta particle velocities solved for with scaling laws from Housen & Holsapple (2011): a tub (or *bowl*) of 641,586 particles representing a portion of the surface of an asteroid that suffers a kinetic impactor strike is embedded into the surface of a 75-m-radius sphere that represents the entirety of the small body; the particles are then ejected (top-right frame). The bottom frame shows the bowl of particles colored blue and the surface of the spherical body rendered translucent (see Fig. 2 for detailed schematics of this setup). A small particle is affixed to the center of the spherical body with a mass $m_{sb}$ that represents the mass of the entire asteroid of bulk density 1,300 kg/m$^3$. Particles evolve in accord with their mutual gravities.

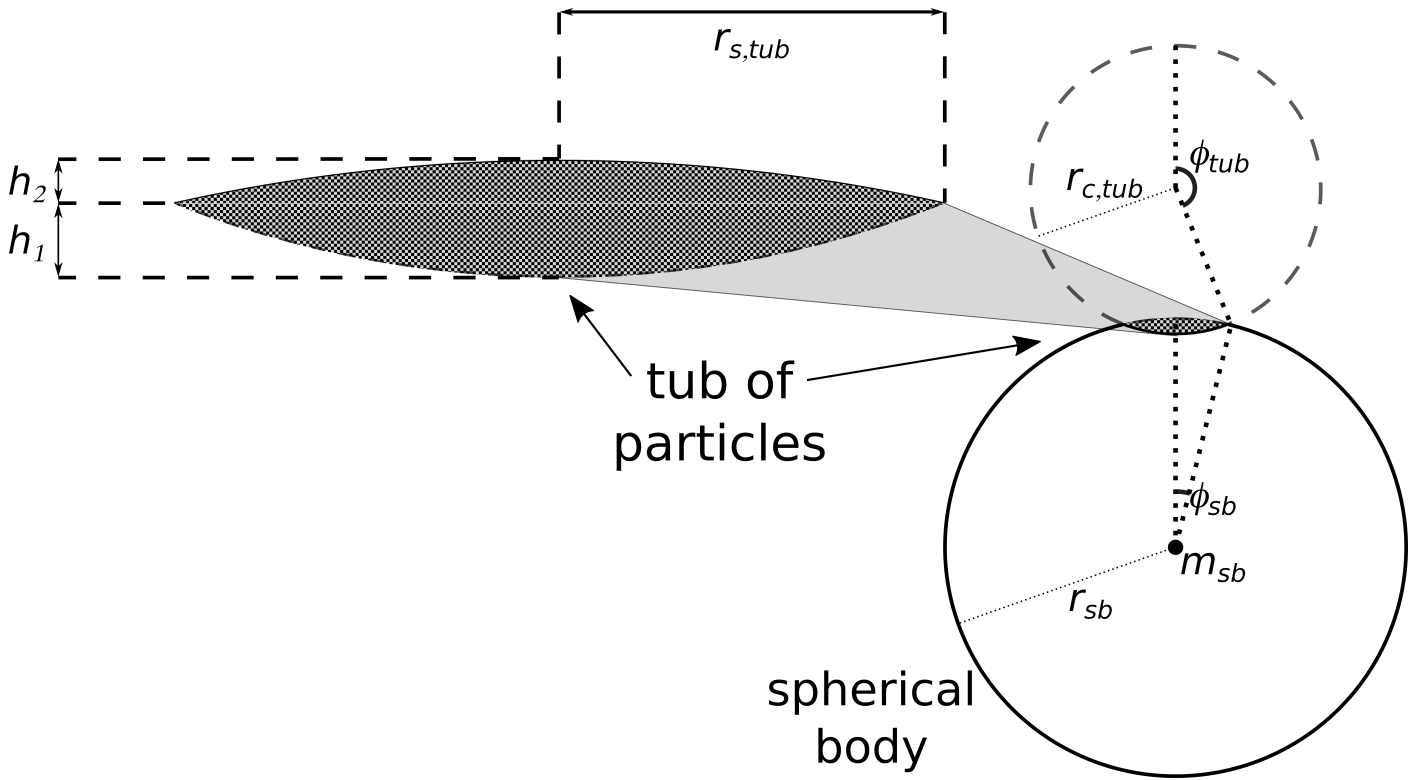

**Figure 2**: The cross-sectional plane containing the centers of both spheres from the configuration described in Section 2.1. Using PKDGRAV (see text), the intersection of these two spheres (shown in detail [left] and in the context of the entire setup [right]) is carefully populated with particles of sizes 14–20 cm. The lower boundary of this intersection, the container *tub*, is defined as a hollow spherical shell with a large opening angle such that $h_1 = R/4$, and $r_{s,tub} = 1.3R$, the height and the surface width, respectively, of the tub. This gives it a curvature radius $r_{c,tub}/R = 701/200$, and an opening angle $\phi_{tub} = \arccos[R/(4 \times r_{c,tub}) - 1] \sim 158.23°$. The spherical body is given an opening angle to match the rim of the tub such that $r_{sb}\sin(\phi_{sb}) = r_{c,tub}\sin(\phi_{tub}) = 1.3R$ (thus $\phi_{sb} \sim 13.533°$). The tub and spherical body are positioned such that their two (identical) rims are flush against each other. The volume of the tub heaped to height $h_2$ with particles rounding out the surface of the spherical body is the sum of the volumes of two spherical caps with radii of curvature $r_{sb}$ and $r_{c,tub}$, and equal surface radii $1.3Rr_{s,tub}$, approximately equal to 2665.1 m$^3$. The small particle that represents the mass of the entire asteroid, $m_{sb}$, is shown affixed to the center of the spherical body.

## 2.2. Using smoothed-particle hydrodynamics to generate initial conditions of ejecta

Below, using a specific example, we describe the results of the SPH simulations of the impact phase, how they are adapted to initial conditions for use in our *N*-Body code, and then we discuss some of the problems encountered thus far with this technique.

### 2.2.1. Outputs from the fragmentation phase

Although the sound wave propagation speed through material can typically be accounted for in our numerical approach based on an *N*-body code using SSDEM for collision handling, an appropriate impact hydrocode is what in general should be used in this regime of fragmentation resulting from a hypervelocity impact (however, there are concerns that arise with the use of SPH to model low-speed ejecta—these are discussed in Section 2.2.3). Thus we use the outputs from Jutzi & Michel (2014), who performed numerical simulations of the impact phase in the framework of the NEOShield project in order to estimate the momentum transfer efficiency of a kinetic impactor over a range of target's properties. They used an SPH impact code specially written to model geologic materials (e.g., Benz & Asphaug, 1995), in which a model was introduced to account for the potential influence of microporosity (Jutzi et al., 2008; 2009). The Tillotson equation of state was used as a tensile fracture model (Benz & Asphaug, 1995), with a standard Drucker–Prager yield criterion for rocky materials that allows shear strength to depend on the confining pressure (Jutzi, 2015).

The initial conditions of the simulations of ejecta fate by PKDGRAV can be given by the outputs of these SPH simulations (see Jutzi & Michel, 2014 for details); these simulations consider a hemispherical domain in the impact region of radius ~17.5 m. The domain used here is bounded during the impact (particles are confined within a finite hemispherical boundary). The number of particles is about 4 million (4,262,972, exactly), which results in a particle mass of 3.15 kg and a particle radius of 0.0754 m, assuming a target bulk density of 1300 kg/m$^3$ and a packing efficiency of about 74% (Hexagonal Close Packing). The projectile's mass and density are 400 kg and 1000 kg/m$^3$, respectively, and its impact direction is vertical along what is defined as the z-axis. The simulations are run for 0.16 sec. Although there does not seem to be high-sensitivity to the end-time of the simulations, investigations are ongoing to better determine the most appropriate handoff-time from SPH to *N*-Body. Once fragmentation is over, the next step is to port SPH outputs into PKDGRAV.

### 2.2.2. Porting SPH output into PKDGRAV

In order to use the output from the SPH-computed impact as initial conditions for an *N*-Body simulation of the ejecta fate, we need to interpret the locations and momenta states of the particles at the end of the SPH simulation. However, because the nature of SPH-coding methodology is such that it treats material as a continuum, in effect the particles do not have a physical radius but rather are characterized by a smoothing length that connects them to the rest of the system. As a result, particles' radii must be assigned. Since SPH contains information on the particle mass, we could assign the radius simply by defining a particle density to conform to the desired bulk density and porosity of the body; nevertheless, it is best to remove all instances of overlap at the start of the *N*-Body simulation to avoid unrealistic mutual repulsion of particle-pairs. Therefore, first, at every location of an SPH particle, we assign an *N*-Body particle with a density equal to the bulk density we assume for the body. Using the masses of each SPH particle, this defines the initial radii of all particles in the system. We next run our *N*-Body code for one step, outputting all instances of particle overlap. For each particle that has overlaps with one or more of its neighbors, the greatest overlap is identified and the particle's radius is shrunk by half this value (e.g., if a particle's greatest overlap is 30% of its radius, the particle radius is shrunk to 85% of its original radius). Our approach entails that particles from areas of high density at the SPH-handoff remain high-density particles for the duration of the simulation. However, we are considering techniques to restore the appropriate densities once particles leave these high-pressure regions.

### 2.2.3. Encountered difficulties using SPH output as initial conditions for the ejecta

Although the procedure as described above does not seem to suffer from major issues, there are in fact some real difficulties in generating the needed data for the study of the gravitational phase of the ejecta from SPH simulations. In particular, the main interest of industries for the design of an observing spacecraft concerns the low-speed ejecta, i.e., those that are ejected at or near the escape speed of the target. These ejecta could trace out convoluted paths within the system, lingering for some time before escaping or falling back onto the asteroid's surface. Given the low gravity, and therefore the low escape speed, of considered targets (of the order of cm/s) as well as the high impact velocity (> 6 km/s, at least five orders of magnitude greater than the escape speed), the computation of the velocities of ejecta get close to the numerical noise of the SPH simulations. In addition, wave reflections at the domain boundary might affect the very low velocity part of the ejecta distribution. Therefore, it is difficult to assess the reliability of the low-speed ejecta output. Moreover, although in the SPH simulations the impact region is initially rigid (strength dominated), in reality, loose particles (regolith) may exist at the surface of an asteroid (even a small-sized one) and the impulse required to loft those particles (provided that they are not stuck by strong cohesive forces) is very small; this is not accounted for in the SPH runs currently performed. In future works, building from our experience and difficulties encountered with these SPH simulations, we plan to perform new SPH simulations using other boundary and surface conditions, pushing the limits of existing software, so that we can access the low-velocity ejecta with greater reliability (and perhaps use other means under study to account for these ejecta when needed).

*2.3. N-Body simulations (post-hypervelocity impact phase)*

Once the particles' post-hypervelocity impact phase positions and ejection velocities are set, either by direct assignment of positions and velocities based on scaling laws (Section 2.1), or by SPH impact simulation (Section 2.2), the *N*-Body simulations to determine the ejecta fate can be performed. For the simplified configuration considered here that includes a homogeneous sphere as the target asteroid, we use a spherical shell boundary to define the asteroid's surface (see Figs. 1–3). For real bodies, greater care must be taken to define the gravitational field and the surfaces (see Section 4).

All particles from the end of the SPH impact phase are then positioned such that those closest to the surface remain just above the surface of spherical wall (Fig. 2). The placement above the wall is to ensure that there are no particles overlapping with the wall (which would entail unrealistic repulsive forces), while still keeping them close to the surface, maintaining the accuracy of the gravity calculations. In the case where we used scaling relations, we were able to preserve the case of a perfect sphere for this baseline simulation by carving out a section of the spherical wall and placing a bowl inside it filled with particles such that the rims of the bowl and of the carved-out section aligned. We also heaped particles on top of the bowl such that the surface of the bulk of particles rounded out the sphere of the 75-m radius body (Fig. 1).

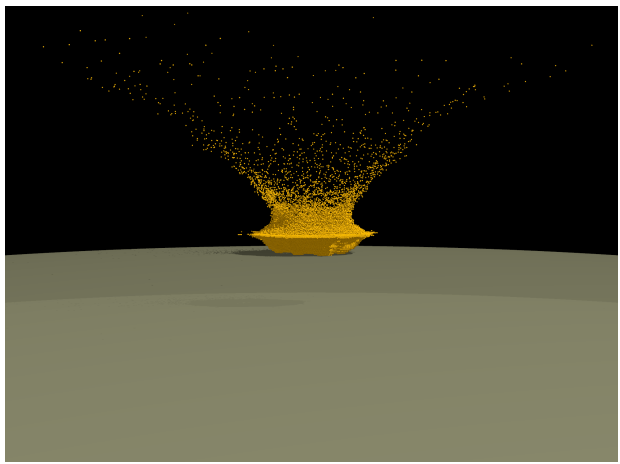

**Figure 3:** Initial positions of those particles moving with +z velocity and with initial angles of trajectory above 10° from the surface after the SPH-computed hypervelocity impact phase for one of the ejecta-evolution simulations involving a 1.5-km asteroid target. We place the ejecta material just above a large sphere that represents the asteroid surface. Although in reality the impact surface should be flush to the surface of the sphere, we take care that the material remains realistically close to the surface.

Particles may interact physically with each other, and so for SPH-derived initial conditions, material parameters are defined to account for the various contact forces/frictions (e.g., static, rolling, sliding, and twisting frictions). For the case using scaling relations, we are in the process of including collisional interactions between particles. Indeed, the expected number of interactions between particles is rather small relative to the total number of particles, as most are on ballistic trajectories and the system is not a particularly dense one. However, collisions could play some role in changing the orbital trajectories of a small fraction of particles, possibly keeping them in the vicinity for more time.

In this study, we do not interest ourselves in the fate of particles that we know will eventually land and remain on the surface of the asteroid. Therefore, as the simulations progress, if a particle impacts the spherical shell (asteroid's surface), we typically make it vanish (delete it from the simulation) when upon impact, which has the advantage of decreasing the number of particles along the way, speeding up the simulation.

For the SPH-derived initial conditions, the procedure to start the next (*N*-Body) phase of simulations, even after the post-hypervelocity-impact-phase particles are successfully ported into PKDGRAV, is not trivial and involves several stages in order to achieve a high level of realism (the procedure based on scaling relations without collisions is less sensitive to the choice of timestep). The first step (for SPH-derived initial conditions) is to perform a portion of the overall simulation using a very small timestep (1 ms) and integrate the system over 100 s (i.e., 99,984 steps). The reasoning here is that because particles are initially moving very fast with very large relative speeds, the stiffness of the spheres has to be large enough to accommodate those relative speeds when they collide, and as per Eq. (4) in Schwartz et al. (2012), the higher stiffness parameter necessitates a smaller timestep. Then, after 100 s, when particles are more dispersed in space and collisions tend to take place occur at lower relative velocities, the stiffness parameter is decreased by a factor of 100, allowing the timestep to be increased to 10 ms. The system is integrated further until 1 ks (90 more ksteps). We continue in this manner, progressively decreasing the particles' stiffness while increasing the timestep with the goal to get to a point where the timestep is large enough such that the system can be evaluated weeks or months after the impact without losing realism. Based upon the speed of simulations thus far carried out, we consider that this happens once a timestep of about 1 sec is reached. This corresponds to a stiffness parameter for very soft particles, where only relatively low-speed collisions can be accurately handled.

All throughout the simulation process, irrespective of the initial conditions chosen, we have various quality controls in place. These include warnings for large particle overlaps, an overall assessment of how many particles are in contact and to what degree of overlap, internal consistency checks (tracking neighbor/contact lists), etc. There are also energy, momentum, and even mass conservation issues to keep track of. For example, particle-particle collisions cause energy loss (due to friction); we do not account for the reaction force on the asteroid when particles collide with it (infinite inertia approximation); also, particles may stick or be deleted upon contact with the asteroid surface. Moreover, we need to note that many SPH material properties that are specific to the SPH phase, such as pressure and material cohesion, are neglected in the *N*-body phase.

*2.4. Current limitations*

There are currently several important limitations in the work that we present here in addition to the significant amount of CPU-time required for simulation. (Using a single-node AMD Opteron 6284-SE machine—32 cores of 2.7 GHz—advancing by one step, a self-gravitating 450,000-particle simulation takes at least 2 seconds, depending on the extent of each particle's neighbor search. During the initial phases of the simulation, timesteps can be as low as 16 μs, which means that one second of simulation time can take 1–2 days to integrate, and it can take several seconds before an increase in timestep is appropriate.)

At present, our *N*-Body modeling does not account for the influence of solar radiation pressure on ejected particles and is therefore adapted to ejecta whose size is large enough as to not be dominated by this process within the first day or so after an impact (> a few cm). A full treatment of solar radiation effects, including the ejecta particles' experiences of solar occultation by a small body in the system or by other ejecta particles, is involved and potentially computationally expensive. However, for particles large enough to not suffer major perturbations on the timescales of these occultations, an approximation of the time-integrated average of the solar radiation pressure can suffice. This is not included here, but is currently being addressed and will be a necessary part of specific kinetic impactor studies, as will be the effects of solar and planetary tides, insofar as they are relevant to the given case.

Other limitations of the model presented here, besides the neglecting of the effects of solar radiation and solar tides, include the use of a single, spherical, non-rotating asteroid; and the use of only spherically-shaped ejecta particles within a certain size range (7–10-cm radii, in the cases of SPH and scaling law initial conditions).

In the realm of granular physics, limiting ourselves to the use of spherical particles is almost always a concern. Although augmenting our code to include non-spherical shapes is an ongoing effort, for the case at hand (ejecta fate, ejecta reaccumulation and granular interactions), frictional effects, including, most significantly in SSDEM, rolling and twisting friction (see Schwartz et al. 2012), should be enough to approximate the effects of irregular shapes on the granular behavior of interest. Concerning our baseline model of the impact target (single, spherical, non-rotating), we intend to apply the methodology described in this work to develop simulations for multiple specific cases of ejecta fate from non-stationary, non-spherical impact targets. The most significant of these is the case of AIDA and the Didymos binary system. However, we reiterate that here we provide a simplified case to elucidate our methodology.

## 3. Results

We have developed and explored each of the methodologies described in Section 2. In addition to following the procedures and performing the simulations, we must also provide ways to describe the resulting ejecta field and its evolution in time. We have developed helpful analysis tools, which we include here as applied to these first, simplified cases. Results are visualized in Figs. 6–8 (scaling relations) and Figs. 9–11 (SPH-derived initial conditions). The first figure in each of the sets (Figs. 6 & 9) shows three cross-sections above the surface of the body through averaged-density space; that is, masses contained in cubic volumes of space (mesh-sizes are specified for each set of plots) are smoothed over these cubic regions. The second figure in these sets (Figs. 7 & 10) shows smoothed energy densities in a similar manner as the first set of figures show smoothed mass densities. The third figure in these sets (Figs. 8 & 11) shows the mass densities within different spatial regions. Here we discuss the results of the cases explored in this study.

### *3.1. Simulation results with initial conditions derived from scaling relations*

Our baseline simulation, the most straightforward to implement, consisted of assigning velocities to particles using experimentally derived scaling relationships (Section 2.1; Figs. 1–2). Figures 6–8, with different ways to represent the outcome using these initial conditions, show density increasing with time at greater distances above the surface in a symmetric manner with little dispersion. Figures 6 & 7, respectively, show the mass and energy densities for cross-sections through the vertical axis for three different heights at six different moments in time after impact. Particles close to the impact site escape quickly to infinity with no significant deviation in their flight path, while slower moving material closer to the nominal rim of the crater are deflected by the gravity of the body; this deflection of slower moving material can be seen in Figs. 6 & 7 by noting the increase in radii of the high-density rings as time progresses. The slowest moving material, which does not make excursions far from the surface, trace close to parabolic trajectories before re-impacting the surface. Between the extremes in ejection speeds, particles cover all spatial locations shallow to 45° from the surface, but at different times (Fig. 4).

In our simulations, the slowly escaping material is affected by the gravity of the body from which it is escaping and by the mutual self-gravity of the ejecta (for the baseline simulation, collisions and solar perturbations are not in effect, but mutual self-gravity between all particles is taken into account). This *gently lofted* material has time to form under-dense and over-dense regions due to the gravitational forces; this can be seen in Fig. 1 (top-right image): the distribution of material in the ejecta curtain is not smooth, and, although this is not apparent from the single snapshot in Fig. 1, it becomes less homogeneous as the time from impact increases. We expect this effect could be exaggerated by the inclusion of dissipative collisions, one of the many additions being built atop this baseline model.

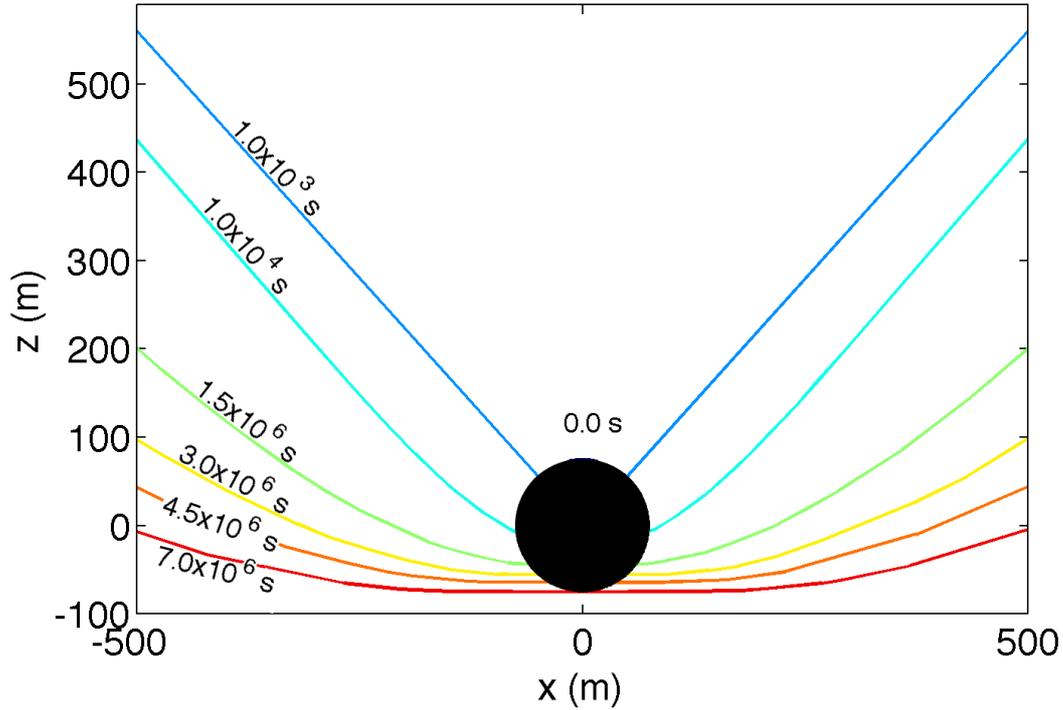

**Figure 4:** The ejecta envelopes at different epochs (azimuthally averaged and projected onto the x-z plane) based on scaling relation-initialized simulation (Section 2.1). The shadow indicates the region of space physically occupied by the asteroid. Slower moving ejecta still close to the asteroid at late epochs have been affected by its gravity. This baseline model assumes a simple, non-rotating spherically shaped body and does not account for solar perturbations (see text), which would clear this vicinity of debris by the later epochs.

*3.2. Simulation results with initial conditions derived from SPH-simulations*

In contrast to the results based on scaling relations, the results using SPH-generated initial conditions show the initial, high-speed ejecta escaping at low angles of inclination, with the angle increasing for ejecta that escape at lower speeds. Figure 5 shows the location of the ejecta as time advances. The top row contains ray-traced images of particle locations just above (within ~25 m of) the surface at four different epochs, while the bottom row contains plots of the azimuthally averaged positions of ejecta at these same epochs, but showing a region of ~6 km. We can see, from the bottom row of Fig. 5, that at 1 ks, ejecta at inclinations below 60° have been cleared from the 6-km vicinity. At later times this region repopulates, albeit sparsely, with some slower ejecta due to gravitational, and possibly collisional effects. By 10 ks (~3 hrs), the ejecta cone within a region of 6 km above the surface has been filled in by slow-moving ejecta (the ejecta is slow-moving relative to the high-speed ejecta, but it should be kept in mind that the majority of these ejecta are still at speeds well above the escape speed of the body). This narrowing of the ejecta cone (Figs. 5, 9–11) is in contrast to Figs. 6 & 7, which, although representing different ejecta time-scales, shows the ejecta cone derived from scaling law relations widening as the integration progresses. The SPH-derived initial conditions also lead to a more dispersed ejecta cone (Figs. 5 & 11) than the thin ejecta cone resulting from the idealized initial conditions derived from scaling relations (Fig. 8). Additional comparisons between the results of this methodology and the one based upon assigning initial velocities by scaling relations are discussed in Section 4.

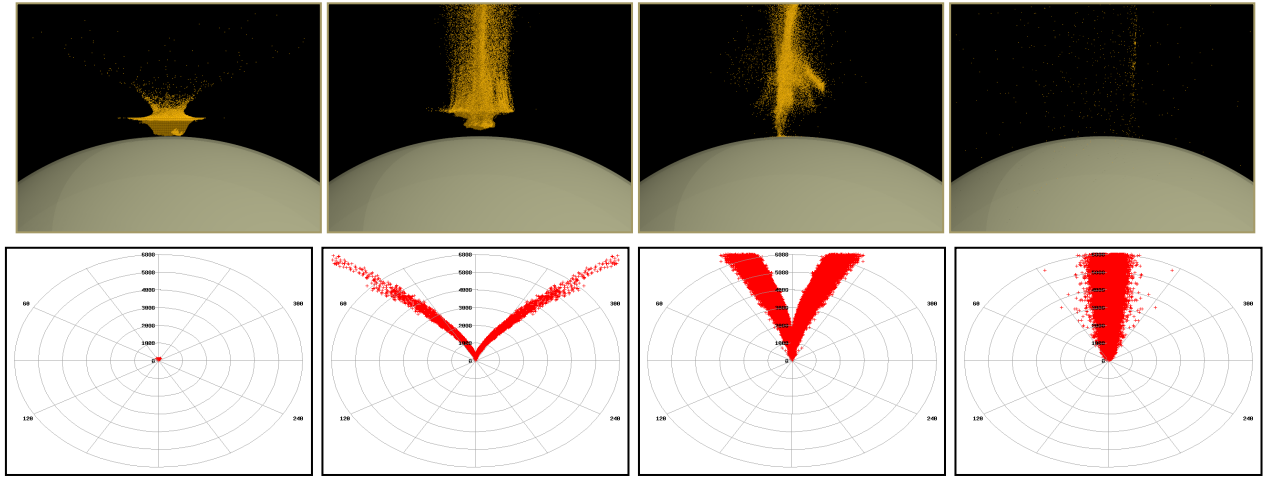

**Figure 5:** Positions of individual ejecta grains after a simulated hypervelocity impact on a 150-m bounded target (notice the curvature of the target compared with the 1.5-km target of Fig. 3). Top row: ray-traced images near the impact point; the images are centered on the point of impact. Bottom row: corresponding polar plots showing particle positions (red points) with concentric circles every 1 km of simulated space; azimuthal positions are not shown and the plots are centered on the asteroid center. In all images, the impact trajectory is downward (z-axis points up). Snapshots progress in time from left to right, with the first frame corresponding to the SPH/ PKDGRAV handoff 0.16 s post-impact; the second frame is taken at 100 s; the third at 1 ks; and the final frame depicted here at 10 ks.

## 4. Analyses and perspectives

We have presented our developing approaches to characterize the vicinity around a small body after an artificial projectile impact. The test case investigated here consists of a non-rotating, spherical body of uniform bulk density 1,300 kg/m$^3$, measuring 150 m across, and impacted normal to the surface at 6 km/s. These approaches all involve the numerical simulation of ejecta using an *N*-Body code to calculate their evolution accounting for their mutual self-gravity. Each of the methodologies includes detailed procedures that determine the initial placements of particles and the arranging of the gravitational environment. Where they differ is in how the initial conditions of the ejecta are generated. The models at the moment do not incorporate the effects of solar radiation or those of solar (or planetary) tides, although we are in the process of adding in these effects.

The two approaches detailed in this study produce results quite different from each other. This is perhaps most apparent when considering the ejection angles of excavated material. Although appropriate for the development of a baseline model (based upon scaling laws), it is clear that assigning a uniform ejection angle of 45° is too simplistic for real scenarios. On the other hand, in the case of SPH, it is also unclear how realistic the result is of slow-moving ejecta lofting off the surface nearly vertically; this phenomenon is not typically observed in impact experiments, however, such experiments have little in common with the microgravity regime studied here. This result does seem to be influenced, at least in part, by the chosen domain and boundary conditions during the SPH simulation: we have examined SPH initial conditions that computed the impact into the same hemispherical region without a boundary, and both cases result in low-speed ejecta launched at high inclination angles, although the angles are less extreme in the unbound case. It could be that in reality much, but not all, of this low-speed material lofting off the surface would be compacted into void spaces, wedged into configurations that would not allow them to escape as individual particles.

Note that however crucial to the design and interpretation of small-body missions such as OSIRIS-REx (NASA), Hayabusa-2 (JAXA), and Rosetta (ESA), we do not have truly reliable experiments that tell us what to expect from impacts into loose regolith in microgravity environments (see the discussion on the Deep Impact mission in Section 1). In the framework of the approach using SPH simulations, we will investigate this further by experimenting with the size of the SPH domain and the boundary conditions of the domain. Since the methodology in Section 2.2 is adaptable to most other software platforms, we are also in the process of using multiple different classes of hydrocodes including different SPH codes to examine this and other ejecta behavior (and, the methodology in Section 2.1 is adaptable to other scaling relations as well). We will also incorporate more realistic ejection angles from scaling relations (still, there is only so much scaling relations can currently provide about this low-speed regime).

Related to the high-inclination ejection angles of low-speed ejecta, we have determined that the filamentary strings seen to develop in the SPH-derived ejecta (Fig. 5) to be highly affected by the tidal forces of the body. Energy-dissipating collisions and/or inter-particle gravity may also help by playing a role, but the effect is greatly diminished when the size of the body is increased from 150 m to 1.5 km.

Additional methods to account for the lofting of regolith material are also being pursued, in part because there is to date no good experimental predictions for particles moving at speeds of order centimeters-per-second for long periods of time under microgravity conditions. Experiments such as these would be needed to tune scaling parameters such as $p$ in Housen & Holsapple (2011), which controls the cutoff at low speeds, or to revisit and redefine these laws with more delicately for crater boundary regions. In the setup of Section 2.1, we began by using $p = 0.3$, as suggested in Housen & Holsapple (2011), but this produced *no* ejecta below the escape speed. Although, their scaling laws do not extend down to anywhere near these speeds, we needed to alter this parameter and to ensure that the size of the tub extended beyond $n_2 R$ in order to produce low-speed ejecta. Since our choice of this parameter does not have experimental justification, we believe that solving for the very low-speed regolith with a granular code is one promising course to take. We are currently working to compare the ejecta to the scaling relations at speeds where both *N*-Body ejecta distributions and scaling relations are justified. We can then, in effect, extend the scaling relations to the distributions of low-speed ejecta, essential for this study on ejecta fate.

Here are a few ongoing improvements: (1) In order to match the shapes and spins of specific bodies and accurately account for gravitational potential, in ongoing studies (to be presented in future work, and not included here) we deviate from the assumption of a homogenous perfect sphere and instead model the body (or bodies, in the case of multiple-asteroid systems such as Didymos) as collections of particles given by an actual shape model. We give these particles solid-body rotation based on NEO spin rate and account for ejecta based on surface interactions and the time-varying gravitational field. (2) The effects of solar radiation and solar tides will be added to the simulations; planetary tides will also be added as needed. (3) We will be including collisional effects into the simulations that use scaling relations to assign velocities: we have just recently calculated an appropriate range of particle stiffness to use in these simulations. (4) Thus far with the *N*-Body-only approach, we have simulated cohesionless regolith, and it might be that some cohesion is called for (we can provide this—see Schwartz et al., 2013). (5) So far we have not seen a need to use HSDEM in place of SSDEM at the late-stage ejecta phases, however, since HSDEM can be integrated faster in situations where collisions are rare, we continue to keep this option in mind. (6) We are working on more ways to adapt the setup from the simplified case studied here to the much more complex case of the Didymos binary system, the target of the proposed NEO-deflection mission, AIDA. Although the case of the binary makes it a challenge, we can take advantage of the assumption that the secondary is tidally locked to the primary. (7) The technique in Section 2.2.2 involves shrinking SPH-particle radii such that there are no overlaps at the start of the *N*-Body simulation. In particularly dense regions, this can lead to the system containing a small amount of unrealistically dense ejecta. We don't see this as a major concern because the most significant effect of having dense particles is that they have smaller cross-sections, and collisions far from the body are rare. Nevertheless, we have developed a strategy (yet to be implemented) to allow these over-dense particles to restore an appropriate radius in a way that conserves angular momentum. (8) Integrating the volumetric distributions of mass and energy (Figs. 6–11) over specified time intervals correlates to the chances of finding hazardous ejecta in these regions during such intervals, thus providing a quantitative sense of the relative safety of these regions. These types of calculations will be performed for more complex, realistic cases in future work.

As one can see, the realistic prediction of trajectories of material ejected as a result of impact cratering events or deflection tests is a complex numerical project. Such an undertaking requires detailed numerical approaches and significant computational time. With so many variables, many of which remain unknown, a perfect assessment of ejecta behavior is simply not possible. We present here the developing stages of this task with much testing yet to be performed, however, we have come up with sophisticated strategies to integrate impact ejecta forward in time given a range of initial conditions and assumptions. We have also developed ways to visualize and assess quantitatively the ejecta trajectories.

The outcome of this study has been to define a protocol to compute and analyze ejecta evolution using as inputs the outcomes of fragmentation processes derived from different approaches. One approach uses analytical scaling laws, which has the benefit of providing straightforward prescriptions defining the initial excavation conditions to compute the *N*-Body evolution. However, these inputs are necessarily highly idealized and extrapolated from experimental (Earth-based) impact regimes very different to those we are investigating. The other approach we have detailed here is to use the outcomes from sophisticated hydrocode simulations. In this case, the particle data describing the initial excavation conditions has the benefit of having been derived from the computed outputs of the fragmentation phase of the given impact. However, difficulties arise since SPH particles are not physical particles and have many parameters that are not straightforward to translate into fully discretized *N*-Body particles. Although we do not provide applications to actual scenarios, the results of this study are to identify and explore different approaches that can provide inputs for our *N*-Body code to compute the ejecta evolution. These inputs can include a wide range of initial conditions based upon both analytical predictions and numerical simulations of outcomes of the fragmentation process. Further, we provide detailed strategies to properly handle these inputs for detailed models of ejecta evolution based on specific impact scenarios and NEO-deflection test contexts.

**Acknowledgements**

This study is performed in the context of the NEOShield project, funded under the *European Commission's FP7 programme* under grant agreement No. 282703, and that of the NEOShield-2 project, which has received funding from the *European Union's Horizon 2020 research and innovation programme* under grant agreement No. 640351. Most of the *N*-Body computation was performed using the Beowulf computing cluster (YORP), run by the Center for Theory and Computation at the University of Maryland's Department of Astronomy. For data visualization, the authors made use of the freeware, multi-platform ray-tracing package, Persistence of Vision Raytracer (POV-RAY).

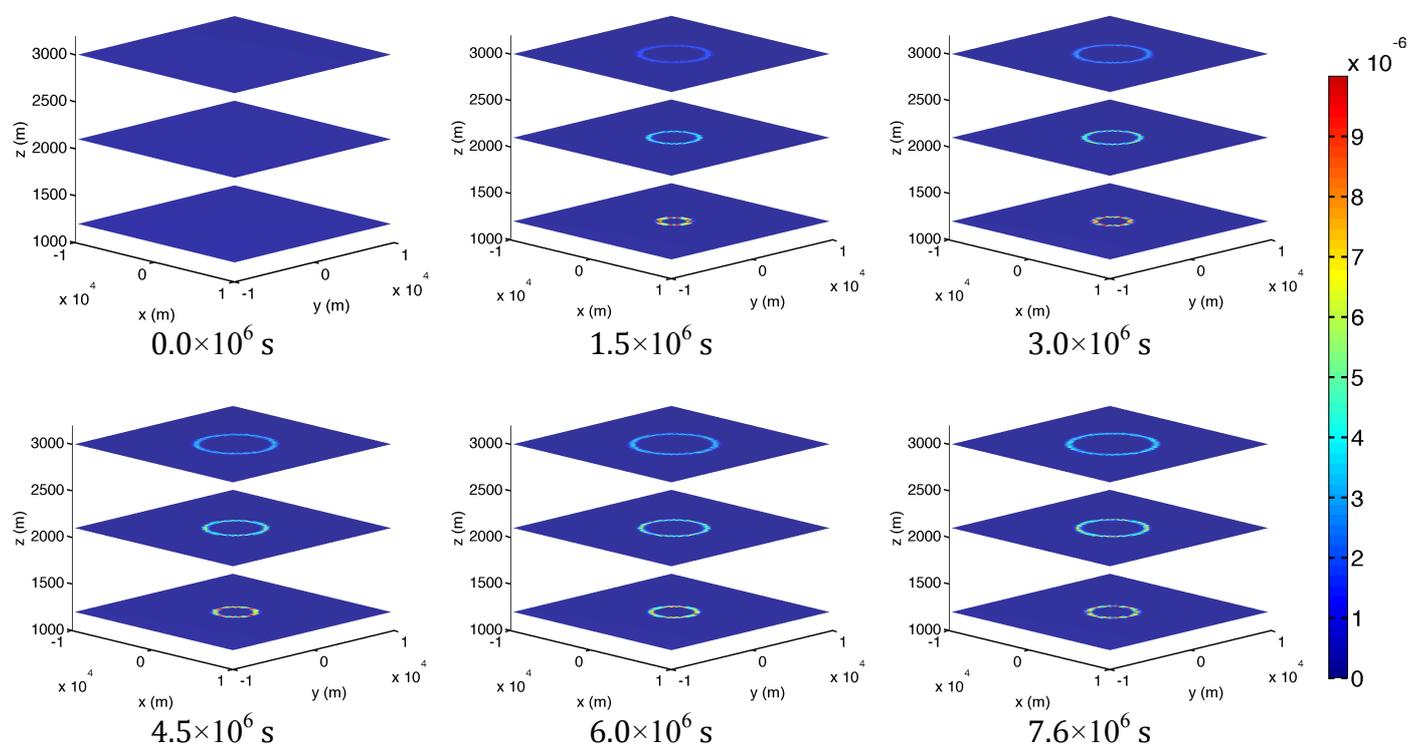

**Figure 6:** Orthogonal-slice planes through volumetric mass density distribution (kg/m$^3$). Mesh width = 0.3$\overline{3}$ km; slices at heights 1.2 km, 2.1 km, and 3.0 km. Based on Scaling Law initialized simulation (Section 2.1). Analysis and visualization techniques like these will be needed to digest the results of less idealized scenarios, including specific cases of actual NEOs.

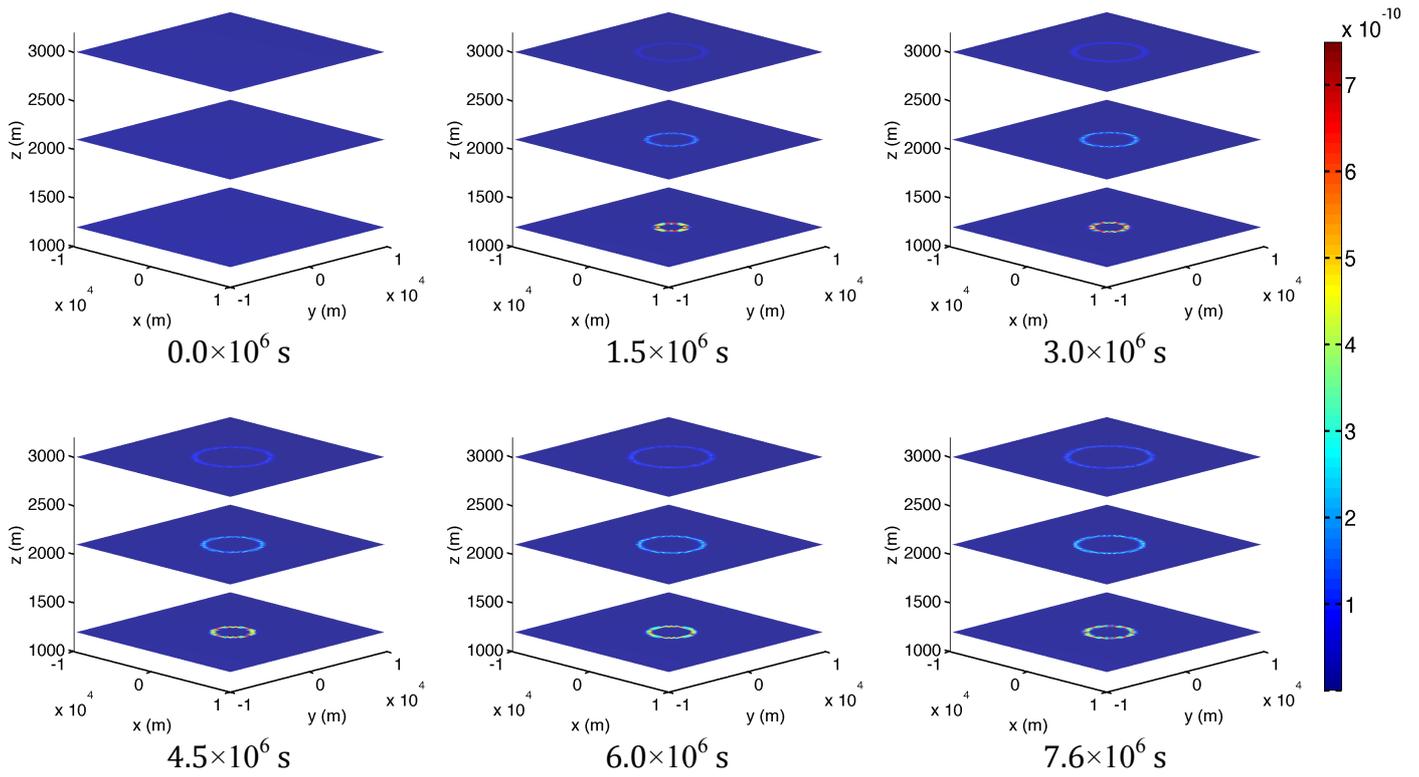

**Figure 7:** Orthogonal-slice planes through volumetric (kinetic) energy density distribution (J/m$^3$). Mesh width = 0.3$\overline{3}$ km; slices at heights 1.2 km, 2.1 km, and 3.0 km. Based on Scaling Law initialized simulation (Section 2.1).

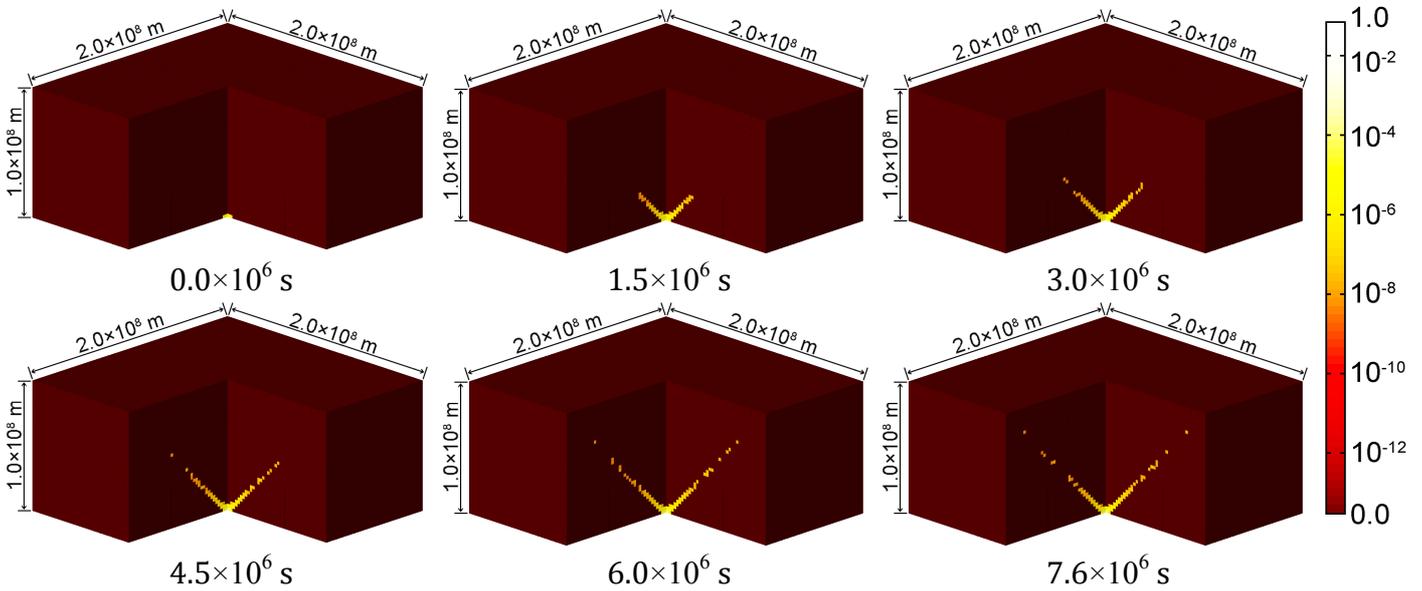

**Figure 8:** Proportional distribution of the ejecta based on Scaling Law initialized simulation (described in Section 2.1). Mesh width = 2.5×10$^3$ km; nonlinear scale is used for the color bar.

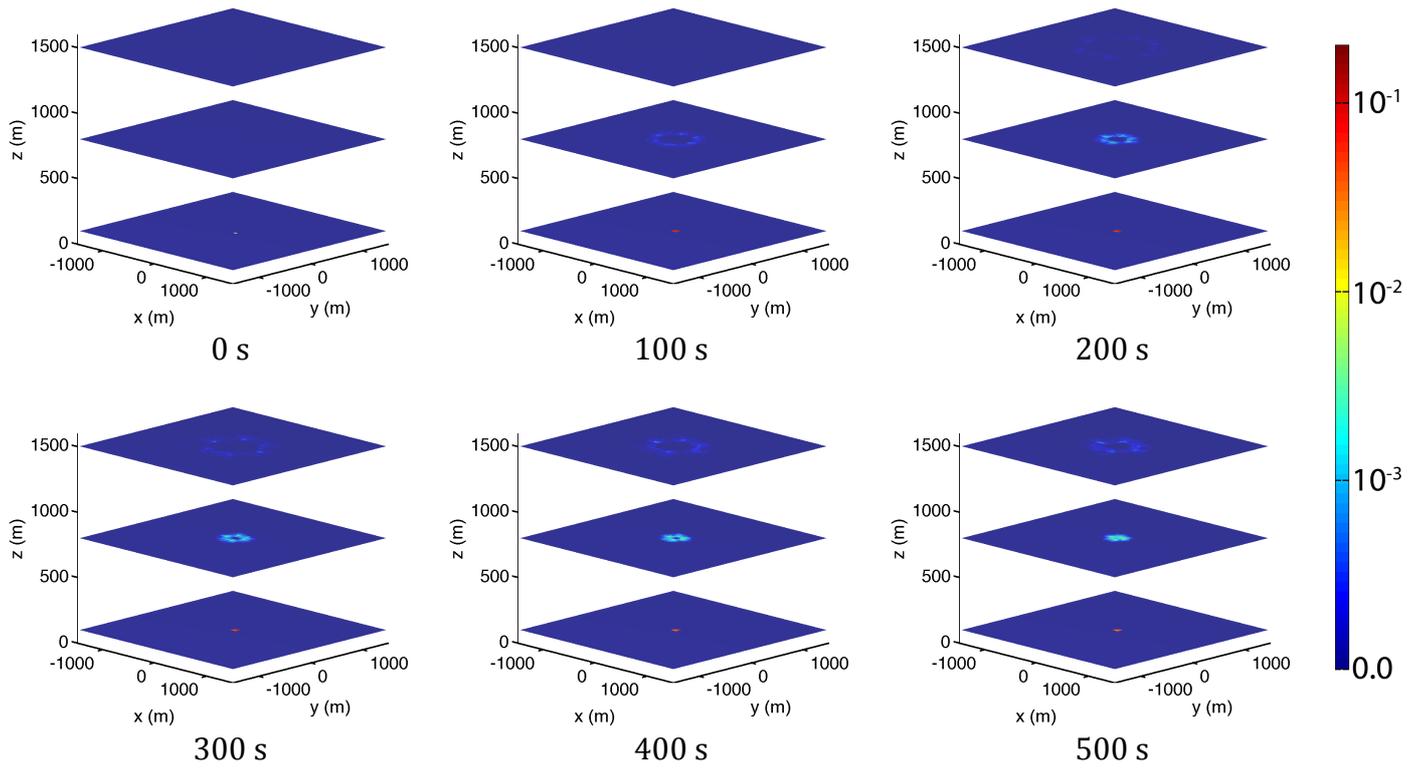

**Figure 9:** Orthogonal slice-planes through volumetric mass density distribution (kg/m$^3$). Mesh width = 50 m; slices at heights 0.1 km, 0.8 km, and 1.5 km. Initial conditions derived from SPH simulation (Section 2.2); nonlinear scale is used for the color bar.

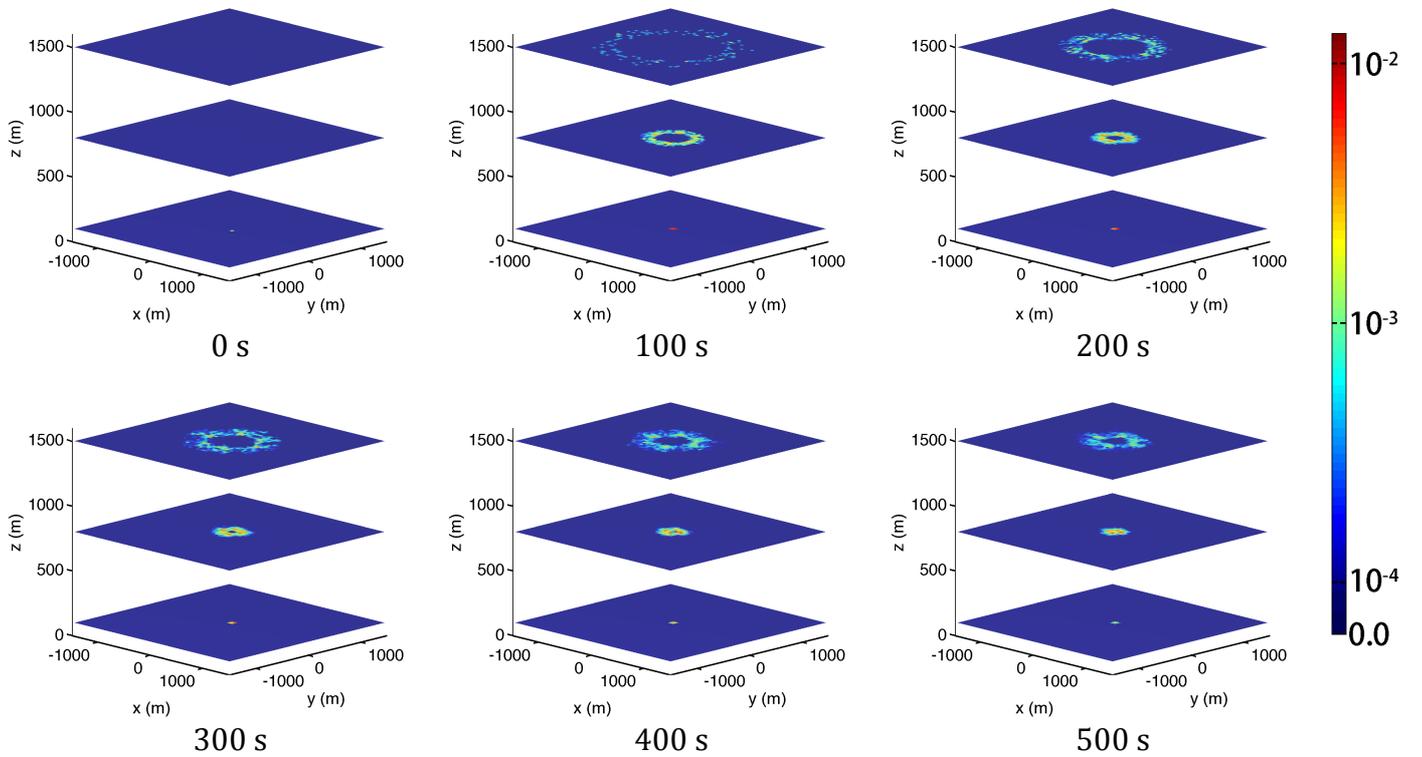

**Figure 10:** Orthogonal slice-planes through volumetric (kinetic) energy density distribution (J/m$^3$). Mesh width = 50 m; slices at heights 0.1 km, 0.8 km, and 1.5 km. Initial conditions derived from SPH simulation (Section 2.2); nonlinear scale is used for the color bar.

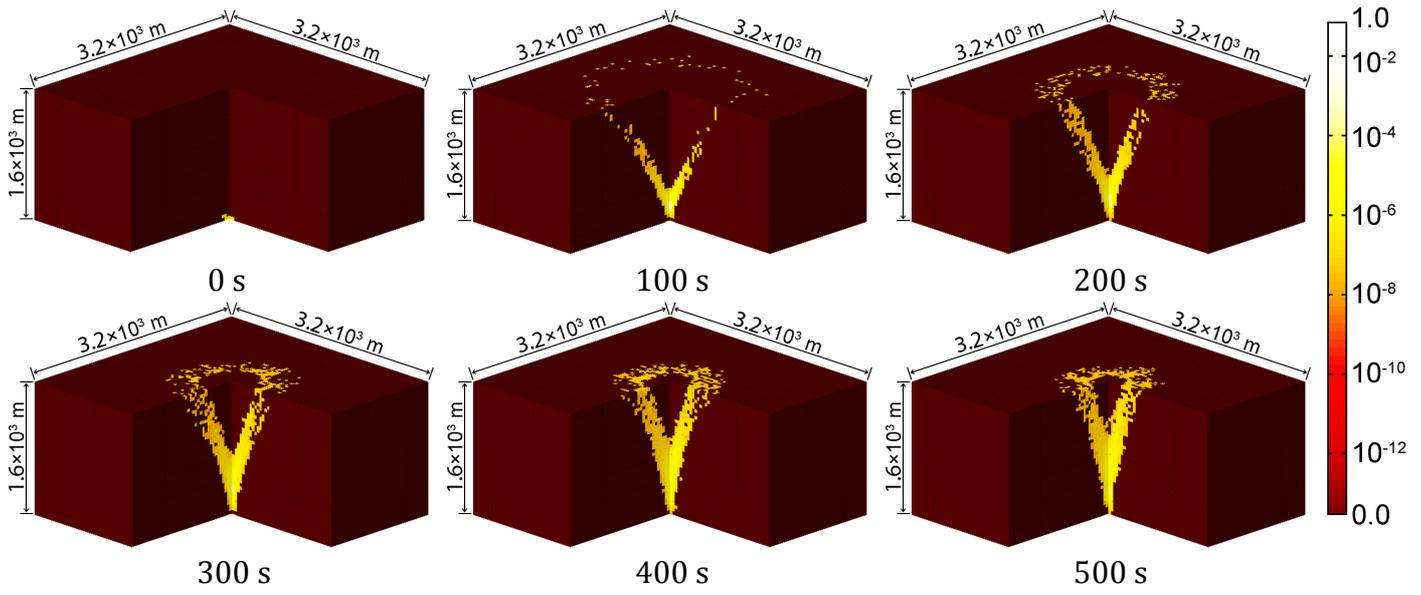

**Figure 11:** Proportional distribution of the ejecta based on initial conditions derived from SPH simulation (Section 2.2). Mesh width = 40 m; nonlinear scale is used for the color bar.